\documentclass[12pt,preprint]{emulateapj}

\shorttitle{WISE MIR-based SFRs}
\shortauthors{LEE ET AL.}

\begin{document}

\title{THE CALIBRATION OF STAR FORMATION RATE INDICATORS
       FOR WISE 22 MICRON SELECTED GALAXIES IN THE SDSS}

\author{Jong Chul Lee$^{1}$, Ho Seong Hwang$^{2}$, and Jongwan Ko$^{1}$}
\affil{$^{1}$ Korea Astronomy and Space Science Institute, Daejeon 305-348, 
              Republic of Korea; jclee@kasi.re.kr\\
       $^{2}$ Smithsonian Astrophysical Observatory, 60 Garden Street, Cambridge, 
              MA 02138, USA\\}

\begin{abstract}
We study star formation rate (SFR) indicators
  for {\it Wide-field Infrared Survey Explorer} ({\it WISE}) 22 \micron{} selected, 
  star-forming galaxies at $0.01 < z < 0.3$ in the Sloan Digital Sky Survey.
Using extinction-corrected 
  H$\alpha$ luminosities and total infrared luminosities
  as reference SFR estimates,
  we calibrate {\it WISE} mid-infrared (MIR) related SFR indicators.
Both 12 and 22 \micron{} monochromatic luminosities 
  correlate well with the reference SFR estimates, but 
  tend to underestimate SFRs of metal-poor galaxies (at lower than solar metallicity),
  consistent with previous studies.
We mitigate this metallicity dependence
  using a linear combination of observed H$\alpha$ and {\it WISE} MIR luminosities
  for SFR estimates.
This combination provides robust SFR measurements
  as Kennicutt et al. (2009) applied to {\it Spitzer} data. 
However, we find that the coefficient $a$ in 
  $L_{\rm H\alpha(obs)} + a~L_{\rm MIR}$ increases with SFR,
  and show that a non-linear combination of observed H$\alpha$ and MIR luminosities
  gives the best SFR estimates with small scatters 
  and with little dependence on physical parameters.
Such a combination of H$\alpha$ and MIR luminosities
  for SFR estimates is first applied to {\it WISE} data.  
We provide several SFR recipes using {\it WISE} data applicable to galaxies
  with 0.1 $\lesssim$ SFR ($M_{\odot}$ yr$^{-1}$) $\lesssim$ 100.
\end{abstract}



\section{INTRODUCTION}

Measuring accurate star formation rates (SFRs) of galaxies 
  is important to understand 
  the formation and evolution of galaxies 
  (see \citealt{ken98} and \citealt{ken12} for a review).
Among many SFR indicators,
  the ultraviolet (UV) continuum and 
  hydrogen recombination emission lines (e.g., H$\alpha$, Pa$\alpha$)
  that are directly related to the bulk energy of young massive stars
  are widely used.
However, SFRs based on UV/optical tracers can be very uncertain
  when galaxies suffer from severe dust extinction that is difficult to correct.
In these dusty galaxies,
  the observation in the infrared (IR), where the dust-reprocessed light emerges, 
  is necessary to measure accurate SFRs.

The monochromatic mid-IR (MIR) luminosities can be useful SFR indicators
  because they are tightly correlated with
  total IR luminosities in normal star-forming galaxies \citep[e.g.,][]{rie09,elb11,got11}.
However, there are several components contributing to the MIR luminosities 
  including the thermal continuum emission from heated small grains,
  polycyclic aromatic hydrocarbon (PAH) features,
  silicate absorption, molecular hydrogen lines, 
  and fine-structure lines \citep[see][]{dra07}.
Because of this complication,
  it is important to examine the reliability of MIR-based SFRs
  in each observed band. 
There have been a number of studies that calibrate the SFR indicators
  based on {\it Spitzer} 8 and 24 \micron{} luminosities \citep[e.g.,][]{wu05,alo06,cal07,zhu08}
  and on {\it AKARI} 9 and 18 \micron{} luminosities \citep{yua11}.
Moreover, the energy balance method that combines (M)IR and
  UV/optical measurements can trace both obscured and unobscured 
  star formation, useful for estimating SFRs 
  of various galaxy populations with small scatters \citep[e.g.,][]{ken09,hao11}.

The new all-sky infrared survey 
  with the {\it Wide-field Infrared Survey Explorer} ({\it WISE}) satellite
  provides photometric data for a large sample of galaxies 
  at 3.4--22 \micron{} with excellent sensitivity \citep[]{wri10}.
There are several studies based on {\it WISE} data
  to investigate the correlations between {\it WISE} MIR luminosities and 
  other SFR indicators \citep{don12, shi12, jar13}.
\citet{don12} and \citet{shi12} use the optical SFR as a reference,
  derived from the Sloan Digital Sky Survey (SDSS; \citealt{yor00}) spectra 
  with the aperture correction method of \citet{bri04}.
This aperture correction method assumes that
  the specific SFR (SFR per unit stellar mass) distribution for a given set of colors 
  inside the fiber is the same as that outside, 
  which can introduce a bias of color-dependent SFR calibration
  \citep[see][]{shi11,xia12}.
On the other hand, \citet{jar13} adopt the IR luminosity as a reference SFR indicator,
  but they use only dozens of galaxies with SFR $\lesssim 4~M_{\sun}$ yr$^{-1}$.
    
In this work, we calibrate the {\it WISE} 12 and 22 \micron{} related SFR indicators
  for a large sample of star-forming galaxies in the local universe.
Using extinction-corrected H$\alpha$ luminosities and total IR luminosities
  as reference SFR indicators,
  we first validate the SFR indicators based on 
  MIR monochromatic luminosities, and
  compare them with previous results.
We then show that the combination of MIR and H$\alpha$ luminosities
  provides better SFR estimates than MIR monochromatic luminosities.
We also suggest several SFR recipes applicable to galaxies with a wide range of SFRs.
The structure of this paper is as follows.
Section 2 describes the observational data and sample selection.
Section 3 explains our calibration results.
We discuss the results and conclude in Sections 4 and 5, respectively.
Throughout, we adopt flat $\Lambda$CDM cosmological parameters with 
  $H_{0}$ = 70 km s$^{-1}$ Mpc$^{-1}$, 
  $\Omega_{\Lambda}$ = 0.7, and $\Omega_{m}$ = 0.3.

\section{DATA AND SAMPLE}

\subsection{Observational Data}

We use a spectroscopic sample of galaxies
  in the SDSS data release 7 \citep{aba09}, covering $\sim$8,000 deg$^2$ of the sky
  and nearly complete to $m_r <$ 17.77 (mag).
We adopt the photometric parameters (e.g., $ugriz$-band magnitudes) of galaxies 
  from the SDSS pipeline \citep{sto02}, and the
  spectroscopic parameters including optical emission line fluxes
  from the MPA/JHU value-added galaxy catalogs \citep[VAGCs;][]{tre04}.

For MIR data, we use the {\it WISE} all-sky survey
  catalog\footnote{http://wise2.ipac.caltech.edu/docs/release/allsky/},
  containing uniform data for over 563 million objects at four IR bands. 
The {\it WISE} 3$\sigma$ sensitivity is estimated to be better 
  than 0.05, 0.07, 0.6 and 3.6 mJy
  at 3.4, 4.6, 12 and 22 \micron{}
  in unconfused regions on the ecliptic plane \citep{wri10}.
We identify {\it WISE} counterparts of the SDSS galaxies
  with a matching tolerance of  3\arcsec{} 
  ($\sim$0.5$\times$FWHM of the {\it WISE}  point spread function at 3.4 \micron).
To avoid contamination by nearby sources within the matching tolerance, 
  we select only unique matches; 
  for a given SDSS object, 
  we choose the WISE object closest to the SDSS object and vice versa.
We focus on galaxies with {\it WISE} 22 \micron{} detection 
  (i.e., signal-to-noise (S/N) $\geqslant$ 3) in the spectroscopic sample. 
All these galaxies have S/Ns $\geqslant$ 3 at 12 \micron. 

To obtain the rest-frame (monochromatic) luminosities 
  at 12 and 22 \micron{} bands (hereafter $L_{\rm W3}$ and $L_{\rm W4}$),
  we compute the $K$-corrections for each galaxy using a set of empirical 
  spectral energy distribution (SED) templates and the fitting code in \citet{ass10}.
Each template spans the wavelength range from 0.03 to 30 \micron{}, and 
  represents an old stellar population, a continuously star-forming galaxy, 
  a starburst galaxy, and an active galactic nucleus (AGN).
We apply this code to the combined photometry of SDSS and {\it WISE} (i.e., 9 data points)
  with varying amounts of reddening and absorption by the intergalactic medium.
We use the Petrosian\footnote{The Petrosian magnitude approximately contains
  the total flux of a galaxy \citep{gra05}.} and 
  point source profile-fitting magnitudes for SDSS and {\it WISE} data, respectively.
The amount of K-corrections in {\it WISE} 12 and 22 \micron{} bands 
 is typically $\lesssim$ 0.1 dex for our sample.

\subsection{Sample Selection}

To construct a reliable sample of star-forming galaxies,
  we first remove AGN-host galaxies in our sample.
Among the 22 \micron{} selected galaxies, 
  we use only galaxies satisfying the selection criteria for pure star-forming galaxies 
  in the emission line ratio diagram of [O{\sc iii}]$\lambda5007$/H$\beta$ versus 
  [N{\sc ii}]$\lambda6584$/H$\alpha$ \citep{kau03b} and S/N $\geqslant$ 3 for each line flux.
Most of these galaxies ($>$ 99.9\%) have {\it WISE} colors of [3.4]$-$[4.6] $<$ 0.8 (mag in Vega),
 indicating that the AGN contamination is negligible in our sample \citep[][]{ste12}.  

We also restrict our analysis to galaxies at 0.01 $< z <$ 0.3.
The upper redshift limit ensures that H$\alpha$ line is comfortably 
  within the SDSS spectral coverage ($\sim$3800--9200 \AA).
The SDSS spectra were taken with 3\arcsec\ diameter fibers. 
Thus the spectra could be dominated by the light from central regions of galaxies.
To avoid this aperture bias, 
  \citet{kew05} recommended using SDSS galaxies at $z >$ 0.04
  to capture $>$ 20\% of the galaxy light.
However, \citet{hop03} demonstrate that their aperture correction
  method works well even for galaxies at $z >$ 0.01 
 (see also figure 13 in \citealt{bri04}).
We also find that using the lower limit of $z>$ 0.01 does not 
  introduce any bias in our results (see next section). 
By changing the lower redshift limit from 0.04 to 0.01, 
  the number of galaxies increases from 90,523 to 105,753. 
This also results in the increase of SFR range for the sample 
  from 1--100 $M_{\sun}$ yr$^{-1}$ to 0.1--100 $M_{\sun}$ yr$^{-1}$. 
Therefore, we can calibrate the SFR indicators
  for a larger number of galaxies and for a wider SFR range.

\section{CALIBRATION OF STAR FORMATION RATE INDICATORS}

\subsection{H$\alpha$ luminosity as a Reference SFR indicator}

We use the H$\alpha$ luminosity of a galaxy as a reference SFR indicator 
  to calibrate the {\it WISE}-based SFR indicators. 
To estimate H$\alpha$ luminosities of SDSS galaxies,
  it is necessary to convert the H$\alpha$ flux measured from a fiber spectrum
  into the one covering the entire galaxy. 
We perform this aperture correction using the difference between $r$-band 
  Petrosian and fiber magnitudes following \citet{hop03}.
This method assumes that the radial profile of line emission is the same as 
  for stellar light, supported by the observational results in \citet{koo01,koo06}.

The observed H$\alpha$ emission suffers from dust extinction from
  both Milky Way and the host galaxy.
The foreground Galactic extinction is corrected
  with the \citet{car89} extinction curve ($R_{\rm V} = 3.1$) and \citet{sch98} maps.
To correct the internal extinction of star-forming galaxies,
  we use the \citet{cal00} extinction curve ($R_{\rm V} = 4.05$) and
  Balmer decrement with the assumption of 
  intrinsic H$\alpha$/H$\beta =$ 2.86 
  \citep[case B recombination for $T_e=10,000$ K and
  $n_e=100$ cm$^{-3}$;][]{ost06}.
If the observed H$\alpha$/H$\beta$ ratio is smaller than 2.86,
  we do not apply this correction.
The emission line fluxes including H$\alpha$ and H$\beta$ 
  are measured after subtracting the stellar population models 
  of Charlot \& Bruzual (2008, in preparation) from the spectra,
  meaning that the stellar absorption in emission lines is properly corrected
  (but see also \citealt{gro12}).

To convert the corrected H$\alpha$ luminosity into a SFR,
  we adopt the relation in \citet{ken98}: 
  SFR$_{\rm H\alpha}~(M_{\sun}$ yr$^{-1}) = 7.9 \times 10^{-42}~L_{\rm H\alpha}$ (ergs s$^{-1}$). 
This relation assumes a Salpeter initial mass function 
  (IMF; mass range at 0.1--100 $M_{\sun}$) and 
  solar abundances with continuous star formation over time scales of 100 Myr.
The SFRs based on the Salpeter IMF are known to be larger than 
  those based on other IMFs such as Kroupa and Chabrier 
  by a factor of 1.4--1.6 \citep[e.g.,][]{cal07,ken09,rie09}.
Therefore, it is necessary to take into account these offsets 
  when one compares the calibration results with different IMFs.
Detailed discussion on the effect of different assumptions can be found 
  in \citet{ken98} and \citet{ken12}.  

\begin{figure*}
\begin{center}
\includegraphics[width=150mm]{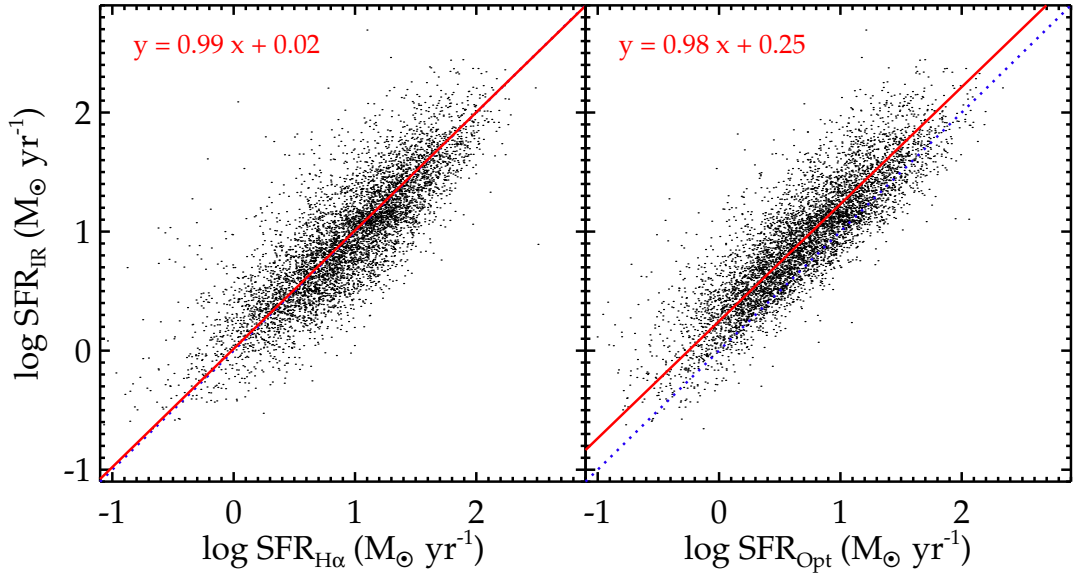} 
\end{center}
\caption{Comparison of SFRs from IR luminosity with
  those from H$\alpha$ luminosities (left) and with
  those from the SDSS MPA catalog (right; \citealt{bri04}).
The red solid lines are the best fits to the data, and
  the blue dotted line are the one-to-one relations.
}\label{fig-base}
\end{figure*}

In Figure \ref{fig-base}, 
  we compare the H$\alpha$-based SFR estimates for our sample galaxies
  with those from total IR (8--1000 \micron) luminosities.
Among 105,753 galaxies in our sample,
  there are 5,995 galaxies with {\it IRAS} 60 \micron{} detection \citep{mos92}.
For these galaxies,
  we compute the total IR luminosities 
  using the SED templates of \citet{cha01}.
We also use 100 \micron{} data for the computation when available
  (see \citealt{hwa10a} for more details).
We convert the IR luminosities into SFRs with the relation of \citet{ken98}: 
  SFR$_{\rm IR}~(M_{\sun}$ yr$^{-1}) = 1.72 \times 10^{-10}~L_{\rm IR}~(L_{\sun})$.

The left panel of Figure \ref{fig-base} shows 
  that SFR$_{\rm H\alpha}$ and SFR$_{\rm IR}$ agree well.
The Calzetti reddening curve seems to work well for the internal extinction correction 
  at least in our sample.
If we use the Cardelli reddening curve instead,
  SFR$_{\rm H\alpha}$ tends to be smaller than SFR$_{\rm IR}$
  for high SFR galaxies (see also figure 3 in \citealt{hwa10a}).

When using the SFRs in the MPA/JHU VAGC derived from the SDSS optical spectra 
  (hereafter SFR$_{\rm Opt}$) rather than SFR$_{\rm H\alpha}$,
  SFR$_{\rm Opt}$ deviates from SFR$_{\rm IR}$ as seen in the right panel.
This offset originates primarily from the aperture correction method of \citet[]{bri04};
  if we compare SFR$_{\rm H\alpha}$ and SFR$_{\rm Opt}$ directly measured from
  the fiber spectra without aperture correction, the two measurements are similar.
  
For the aperture correction of SFR$_{\rm Opt}$, Brinchmann et al. 
  assume that the specific SFR can be estimated from galaxy colors.
However, red galaxies show a very wide range of specific SFRs
  because of the degeneracy between age, metallicity, and extinction \citep[see][]{bri04}.
\citet[]{xia12} indeed show that, in a given set of colors, the specific SFR inside the fiber
  increases with Balmer decrement.
This implies that the color-dependent aperture correction method can result in 
  the underestimation of the specific SFRs outside galaxies,
  in particular, for dusty galaxies.
On the other hand, \citet[]{sal07} found that
  the UV-based SFRs agree well with SFR$_{\rm Opt}$
  for the sample of {\it GALEX} selected SDSS galaxies.
This can suggest that the aperture correction of Brinchmann et al.
  introduces no bias at least for less dusty galaxies.
Similarly, if we compare SFR$_{\rm Opt}$ with our SFR$_{\rm H\alpha}$ 
  for the entire sample of 
  star-forming galaxies regardless of (M)IR detection, 
  the systematic difference is negligible.
Therefore, these results suggest that the effect
  of different aperture corrections is significant
  only for dusty galaxies; the aperture correction 
  of \citet[]{hop03}  based on the simple scaling method
  works well at least in our sample.
  
\subsection{SFR Indicators based on Monochromatic MIR luminosities}

\subsubsection{SFR calibration for 12 and 22 \micron{} luminosities}

\begin{figure*}
\begin{center}
\includegraphics[width=150mm]{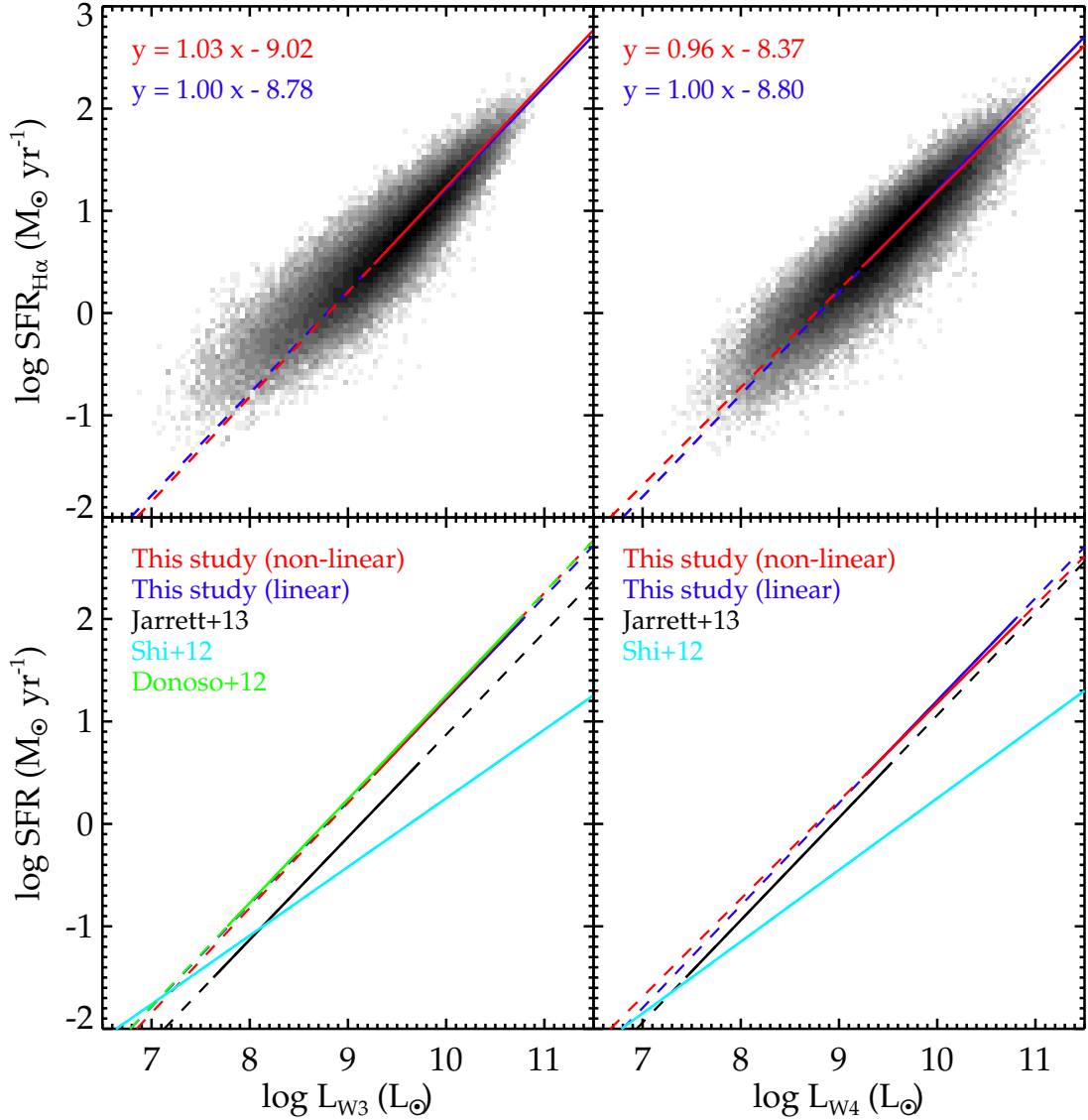}
\end{center}
\caption{{\it Top:} H$\alpha$-based SFRs vs. 
12 \micron{} (left) and 22 \micron{} (right) luminosities.
The data points are indicated as gray-scale density maps for better visibility.
The red solid lines are the best-fits to the galaxies with SFR $>$ 3 $M_{\sun}$ yr$^{-1}$.
The blue solid lines are the same, but by fixing the slope to be unity.
The dashed lines are extensions of the solid lines.
{\it Bottom:} SFR calibrations based on 12 \micron{} (left) and 22 \micron{} (right) luminosities
  of \citet[green line]{don12}, \citet[cyan line]{shi12}, \citet[black line]{jar13}, 
  and this study (red and blue lines).
The solid lines indicate the SFR ranges used for deriving the relations (see Table 1), 
  and the dashed lines extend these relations.
}\label{fig-mono}
\end{figure*}

In the top panels of Figure \ref{fig-mono}, 
  we plot ${\rm SFR_{H\alpha}}$ as a function of $L_{\rm W3}$ and $L_{\rm W4}$.
Both panels show good correlations,
  but the slope of the relation between ${\rm SFR_{H\alpha}}$ and $L_{\rm W3}$ 
  appears to change.  
We use the bisector method \citep{iso90} to fit the galaxies 
  with SFR $> 3~M_{\sun}$ yr$^{-1}$,
  and obtain the following relations (red solid lines):
\begin{equation}
{\rm SFR_{W3}}~(M_{\sun}~{\rm yr}^{-1}) = \\ 
(9.54\pm 0.44) \times 10^{-10}~{L_{\rm W3}}^{1.03\pm 0.01}~(L_{\sun}),
\end{equation}
\begin{equation}
{\rm SFR_{W4}}~(M_{\sun}~{\rm yr}^{-1}) = \\
(4.25\pm 0.20) \times 10^{-9}~{L_{\rm W4}}^{0.96\pm 0.01}~(L_{\sun}).
\end{equation}
We choose the fitting range of SFR $> 3~M_{\sun}$ yr$^{-1}$ where the slope converges
  within the fitting error.
If we fix the slope to be unity,
  the resulting relations (blue solid lines) are
  ${\rm SFR_{W3}}~(M_{\sun}$ yr$^{-1}) = (1.64\pm 0.11) \times 10^{-9}~{L_{\rm W3}}~(L_{\sun})$ and
  ${\rm SFR_{W4}}~(M_{\sun}$ yr$^{-1}) = (1.59\pm 0.11) \times 10^{-9}~{L_{\rm W4}}~(L_{\sun})$.

The top panels suggest that  
  12 and 22 \micron{} monochromatic luminosities
  can be good SFR indicators,
  considering the tightness of the correlations 
  (Spearman rank correlation coefficients = $\sim$0.8 and 
  dispersions of the fitting residuals = $\sim$0.2 dex).
The scatters in the correlations are not fully explained by 
  the expectations from measurement errors 
  (typically 0.01, 0.02, and 0.07 dex 
  for H$\alpha$ fluxes, 12 and 22 micron flux densities, respectively).
These scatters mainly result from 
  the uncertainties in the corrections for H$\alpha$ luminosities
(i.e., $\sim$0.18 and $\sim$0.08 dex for extinction and aperture corrections, respectively).
Note also that the relation for 12 \micron{} luminosity is meaningful only 
  for galaxies with SFR $\gtrsim 1~M_{\sun}$ yr$^{-1}$
  because galaxies with SFR $\lesssim 1~M_{\sun}$ yr$^{-1}$ deviate 
  more than 1$\sigma$ from the relation for high SFR galaxies.

We summarize the relations between SFRs and
  12/22 \micron{} luminosities in Table \ref{table1}
  together with those in the literature at similar wavelengths.
To directly compare our results with those in the literature,
  we plot several relations based only on {\it WISE} data in the bottom panels 
  of Figure \ref{fig-mono}.
The figure shows that the results in this study and in \citet{don12} 
  are in excellent agreement.
The results of \citet{jar13} show small offsets from ours,
  but these are not statistically significant. 
However, the results of \citet{shi12} are clearly deviated from 
  other relations, especially in high SFR galaxies.
The exact cause for this difference is not fully understood. 
However, we suspect that the offset mainly results from the difference 
  in computing the MIR luminosities. 
It is because the MIR luminosity range 
  in their sample is much larger than for the samples 
  in this study and in \citet{don12}
  even though all the studies use similar {\it WISE} selected SDSS galaxies.
For example, it is not clearly explained in their paper 
whether they perform the K-corrections to compute the rest-frame MIR luminosities.
  
Our calibration of SFR indicators based on 
  {\it WISE} MIR luminosities is consistent with 
  those in \citet{don12}  and \citet{jar13}.
However,
  our calibration is based on a large sample of galaxies with a wide SFR range,
  and we provide the relations for both 12 and 22 $\mu$m  luminosities, 
    suggesting that our results can supplant previous results.
Moreover, thanks to our large sample,
  we further show that
  the SFRs based on MIR luminosities
  suffer from metallicity bias in the next section.
    
\subsubsection{Dependence of SFR calibration on physical parameters}

To study what makes the slope for the relation between
  ${\rm SFR_{H\alpha}}$ and $L_{\rm W3}$ change,
  we plot the ratio of 
  SFR$_{\rm W3}$ (Equation 1) to SFR$_{\rm H\alpha}$ 
  as a function of several physical parameters in Figure \ref{fig-res1}.
These parameters include (gas-phase) metallicity, stellar mass, mean stellar age, and 
  the amount of dust extinction.
We use the oxygen abundance (12$+$log(O/H) based on optical nebular lines; \citealt{tre04}),
  stellar mass (M$_{\rm star}$ from the SED fit of the SDSS photometry; see also \citealt{kau03a}),
  and light-weighted age (from the stellar absorption features; \citealt{gal05}) 
  in the SDSS MPA/JHU DR7 VAGCs.
The UV continuum slope ($\beta$, defined as $f_{\lambda}\varpropto\lambda^{\beta}$)
  is a good proxy for the amount of dust extinction,
  similar to Balmer decrement \citep[e.g.,][]{meu99}.
However, the UV continuum slope 
  can also be substantially influenced by stellar population age \citep[]{mao12}.
Following \citet{ove11}, we compute the UV continuum slope ($\beta_{\rm GALEX}$)
  from the difference between far- and near-UV magnitudes
  in the {\it GALEX} database\footnote{We use {\it GALEX} general release 6 that 
  provides the cross-matched table against SDSS DR7 (http://galex.stsci.edu/GR6).}.
  
The top left panel shows a strong dependence of SFR$_{\rm W3}$/SFR$_{\rm H\alpha}$
  on metallicity; metal-poor galaxies have much lower SFR$_{\rm W3}$ 
  than SFR$_{\rm H\alpha}$.
This probably results from low dust-to-gas ratios of metal-poor galaxies, and
  thus they are inefficient in reprocessing of UV light by dust 
  (e.g., \citealt{sch09,hwa12}; see also the discussion in Section \ref{limit}).
Metallicity also plays a role in the abundance of PAH molecules relative to 
  the total dust content \citep[e.g.,][]{ros06,mar10};
  the metallicity effect is more prominent 
  in the IR bands such as {\it WISE} 12 \micron{} containing strong PAH features.
However, the physical mechanisms for the correlation between the metallicity 
  and the PAH feature are still inconclusive. 
These could be due to the delayed production of PAHs in low-metallicity 
  galaxies or to PAH destruction mechanisms in harder radiation fields
  of low-metallicity environments \citep[see][]{vei09,cal11}.

In the top right panel, 
  SFR$_{\rm W3}$/SFR$_{\rm H\alpha}$ also changes significantly 
  with stellar mass.
However, when we use only galaxies with a narrow range of metallicities,
  the stellar mass dependence disappears.
This suggests that the dependence of SFR$_{\rm W3}$/SFR$_{\rm H\alpha}$
  on stellar mass originates simply from 
  the well known mass-metallicity relation \citep[e.g.,][]{tre04,zah12}.

The bottom left panel shows that
  SFR$_{\rm W3}$/SFR$_{\rm H\alpha}$ does not depend on mean stellar age.
The MIR emission at 12 \micron{} could be attributed to the circumstellar dust 
  around evolved stars in the asymptotic giant branch (AGB) \citep[e.g.,][]{pio03,ko12,hwa12}.
The AGB dust emission decreases with increasing age 
  but remains for several Gyrs.
Therefore, in galaxies dominated by old stellar populations,
  the MIR emission from the AGB dust could be as important as
  the dust emission related to the current star formation.
However, this figure suggests that the contribution from the AGB dust
  is insignificant in dusty, star-forming galaxies.
    
The bottom right panel shows that
  there is no significant dependence of SFR$_{\rm W3}$/SFR$_{\rm H\alpha}$ 
  on $\beta_{\rm GALEX}$.
On the other hand, if we replace $\beta_{\rm GALEX}$ with Balmer decrement,
  the SFR ratio decreases systematically by $\sim$0.2 dex, 
  consistent with previous studies \citep[e.g.,][]{ken09,xia12}.
Because $\beta_{\rm GALEX}$ and Balmer decrement are related to the different parts of
  dust extinction (i.e., star vs. gas),
  it would be interesting to investigate this difference 
  with a careful modeling in future studies.

\subsection{SFR Indicators based on the Combination of H$\alpha$ and MIR Luminosities}

As shown in the previous section, 
  the MIR-based SFRs can be uncertain in galaxies 
  where dust reprocesses only a small fraction of light of young stars 
  such as metal-poor galaxies.
In this section, we mitigate this problem
  by combining MIR luminosities and observed 
  (i.e., extinction-uncorrected) H$\alpha$ luminosities \citep[see][]{ken09}.
When we combine $L_{\rm H\alpha(obs)}$ and $L_{\rm MIR}$
  as $L_{\rm H\alpha(obs)} + a~L_{\rm MIR}$,
  the combination coefficient $a$ is determined
  from the ratio of H$\alpha$ luminosity difference 
  before/after extinction correction to MIR luminosity,
  as shown in the top panels of Figure \ref{fig-comb}.
By taking median values of the ratios (horizontal dotted lines), 
  we determine the coefficient $a$ 
  for H$\alpha+W3$ and H$\alpha+W4$
  as $0.036\pm0.001$ and $0.034\pm0.001$, respectively.

As a sanity check,
  we plot the extinction-corrected H$\alpha$ luminosity
  versus the combination of observed H$\alpha$ and MIR luminosities 
  using the coefficients above in the bottom panels.
As expected, the two measurements show a good correlation\footnote{
  This is probably because the ordinate and abscissa are not completely independent.
  However, these correlations remain even if we use the total IR luminosity
  as a reference SFR indicator (see Section \ref{lirsfr}).}.
The most interesting feature in these panels 
  is that the data are tilted from the one-to-one relations (dotted lines),
  suggesting a systematic variation of the combination coefficients. 

This variation of the combination coefficients is already illustrated in the top panels
  as red curves (sliding medians); the coefficients increase
  from 0.01 to 0.05 at 6.5 $\lesssim$ log $L_{\rm H\alpha(corr)}~(L_{\sun})~\lesssim$ 9.5
  (i.e., 0.1 $\lesssim$ SFR ($M_{\odot}$ yr$^{-1}$) $\lesssim$ 100).
The need of high coefficients for high SFR galaxies 
  is consistent with the results in \citet{zhu08} and \citet{cal10}.
To take into account this variation of the coefficient $a$,
  we fit the data in the bottom panels
  and obtain the following relations (red solid lines):
\begin{equation}\label{eq-haw3}
{\rm SFR_{H\alpha+W3}}~(M_{\sun}~{\rm yr}^{-1}) = 
(7.67\pm 0.35) \times 10^{-9}~{[L_{\rm H\alpha(obs)}+0.036~L_{\rm W3}]}^{1.07\pm 0.01}~(L_{\sun}),
\end{equation}
\begin{equation}\label{eq-haw4}
{\rm SFR_{H\alpha+W4}}~(M_{\sun}~{\rm yr}^{-1}) = 
(9.12\pm 0.42) \times 10^{-9}~{[L_{\rm H\alpha(obs)}+0.034~L_{\rm W4}]}^{1.06\pm 0.01}~(L_{\sun}).
\end{equation}
This method is equivalent to the one using variable $a$ 
  (that would be expressed as a function of $L_{\rm H\alpha(obs)}$ and/or $L_{\rm MIR}$) 
  in the linear combination.
We summarize the results based on the combination 
  of H$\alpha$ and MIR luminosities in Table \ref{table2}.
For comparison, we also list the results based on 
  the combination of H$\alpha$ and {\it Spitzer} 8/24 $\mu$m luminosities
  in previous studies.

\begin{figure*}
\begin{center}
\includegraphics[width=150mm]{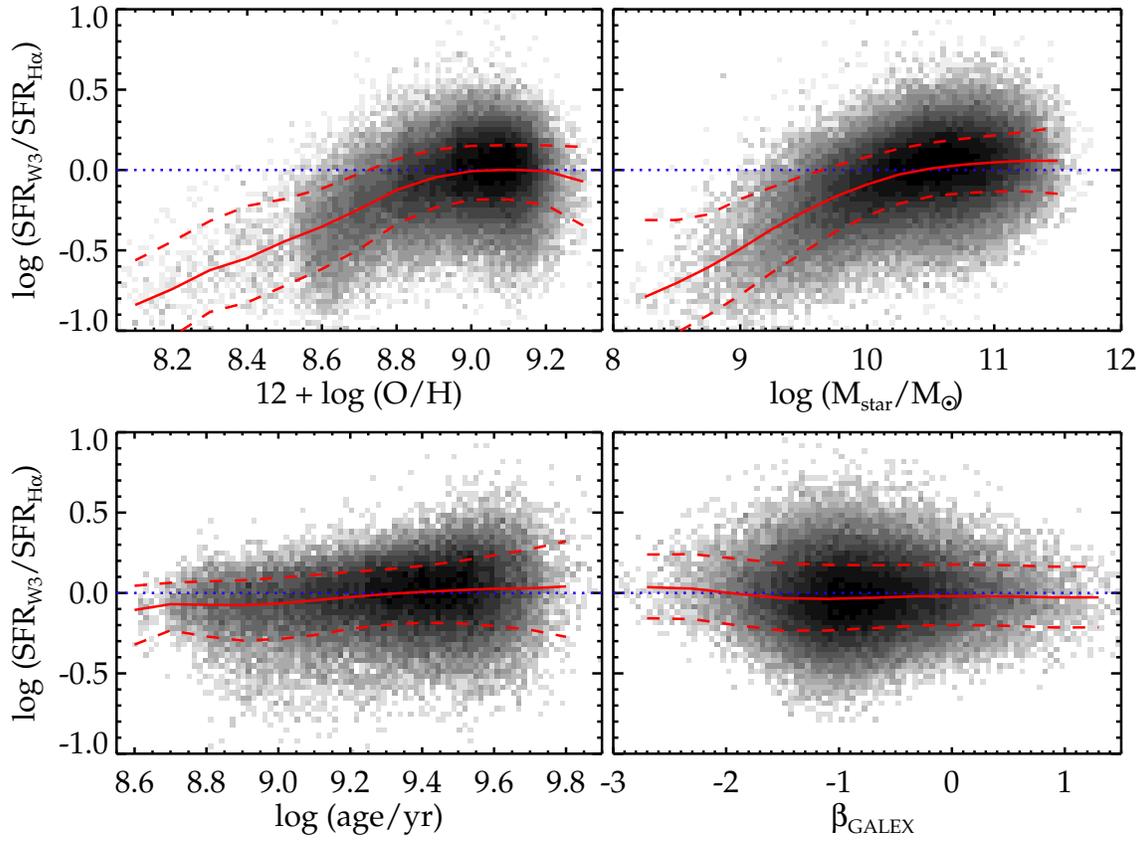}
\end{center}
\caption{Ratio between SFRs derived from 12 \micron{} and 
  from H$\alpha$ luminosities
  as a function of oxygen abundance (top left), stellar mass (top right), 
  mean stellar age (bottom left), and UV continuum slope (bottom right).
The red solid lines represent sliding medians, and
  the red dashed lines enclose 68\% (1$\sigma$) of the galaxies.
The horizontal dotted lines indicate the ratio of unity.
}\label{fig-res1}
\end{figure*}

\begin{figure*}
\begin{center}
\includegraphics[width=150mm]{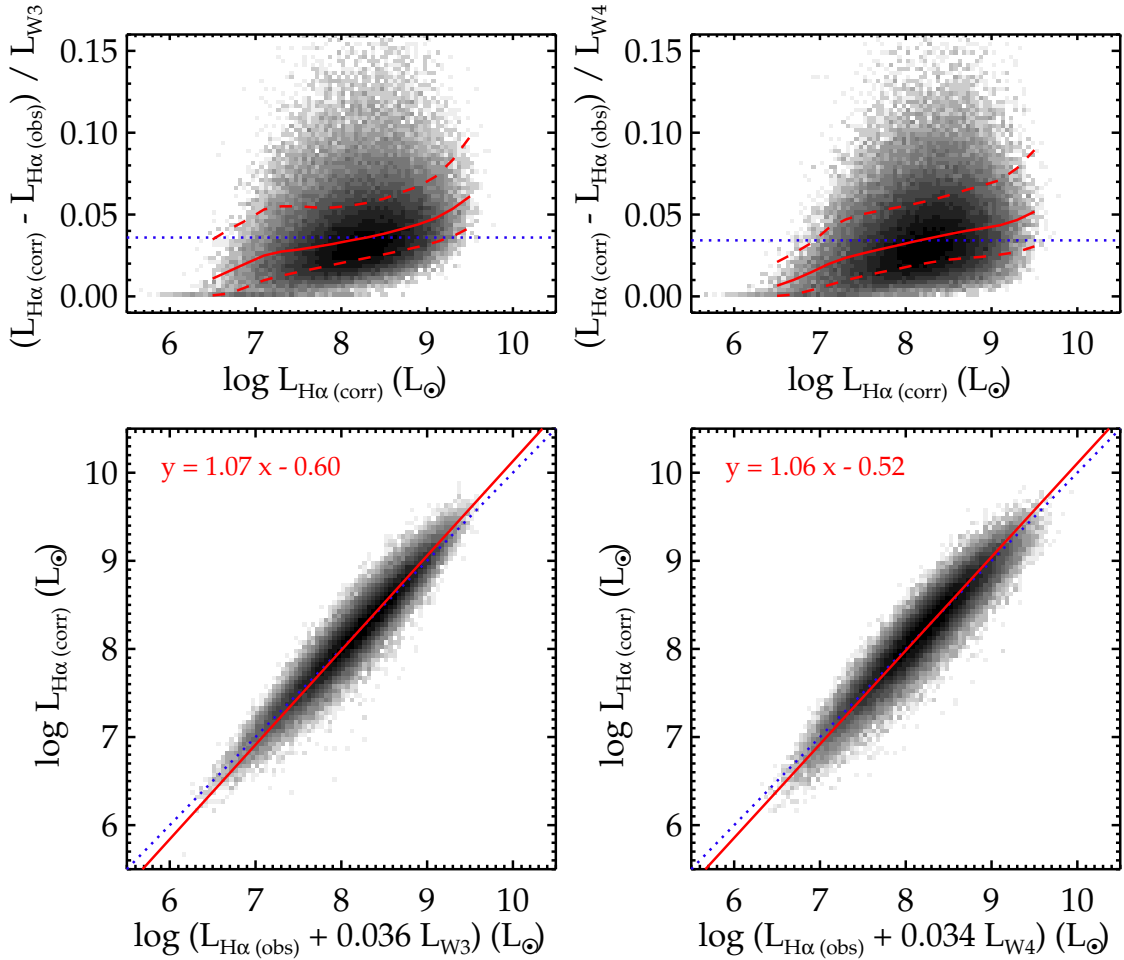}
\end{center}
\caption{{\it Top:} Ratio of H$\alpha$ luminosity difference before/after extinction 
correction to MIR luminosity as a function of extinction-corrected H$\alpha$ luminosity
  (left: 12 \micron, right: 22 \micron).
The blue dotted line is the median ratio in each panel,
  and the red solid line is the sliding median as a function 
  of extinction-corrected H$\alpha$ luminosity.
The red dashed lines enclose 68\% (1$\sigma$) of the galaxies.
{\it Bottom:} Extinction-corrected H$\alpha$ luminosity vs.
combination of observed H$\alpha$ and MIR luminosities 
  (left: 12 \micron, right: 22 \micron).
The red solid line is the best fit to the data, and 
  the blue dotted line is the one-to-one relation.
}\label{fig-comb}
\end{figure*}

\begin{figure*}
\begin{center}
\includegraphics[width=150mm]{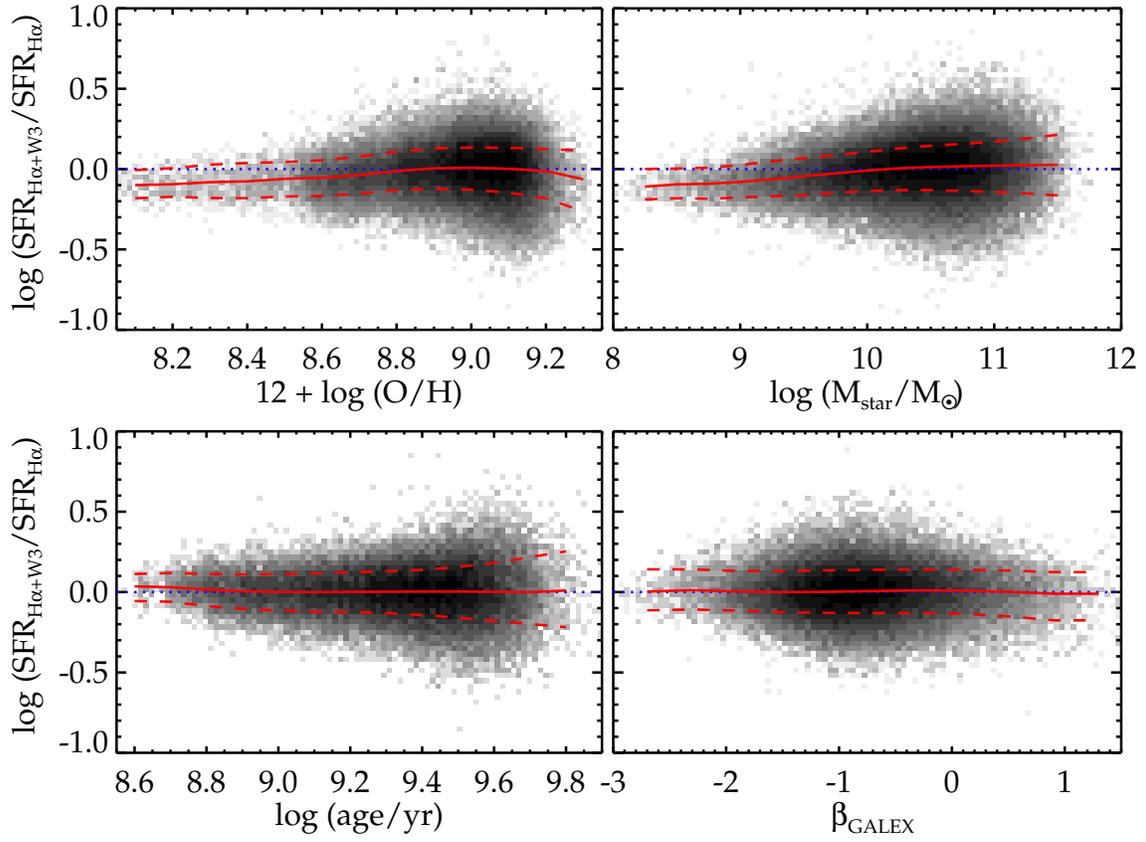} 
\end{center}
\caption{Same as Figure 3, but for the ratio between SFRs
  derived from the combination of observed H$\alpha$ and 12 \micron{} luminosities and
  from the extinction-corrected H$\alpha$ luminosity.
}\label{fig-res2}
\end{figure*}

We then re-examine the dependence of the ratio between 
  SFR$_{\rm H\alpha+W3}$ (Equation 3) and SFR$_{\rm H\alpha}$
  on several physical parameters in Figure \ref{fig-res2}.
The ratio of SFR$_{\rm H\alpha+W3}$/SFR$_{\rm H\alpha}$ depends 
  very weakly on the metallicity (top left panel), 
  different from SFR$_{\rm W3}$/SFR$_{\rm H\alpha}$ (top left panel in Figure \ref{fig-res1}).
The scatter is also small.
The median values for the ratio (red solid curve) 
  above the solar metallicity (12$+$log(O/H)=8.69; \citealt{asp09}) are close to unity, 
  as expected from the assumption for the SFR conversion relation in \citet{ken98}.
At lower metallicities, a small offset seems to exist ($\lesssim$ 0.1 dex),
  but the ratio is still consistent with unity within the uncertainty.
The dependence of SFR$_{\rm H\alpha+W3}$/SFR$_{\rm H\alpha}$
  on stellar mass is again very week (top right panel), similar to the case of metallicity.
The bottom panels show that 
  the dependence of SFR$_{\rm H\alpha+W3}$/SFR$_{\rm H\alpha}$
  on mean stellar age and on UV slope are negligible.
  
The results based on 22 \micron{} luminosities are similar to 
  those in Figures \ref{fig-res1} and \ref{fig-res2} (not shown here),
  but their dependence on metallicity and on stellar mass
  are much weaker than for 12 \micron.

\section{DISCUSSION}

\subsection{Total Infrared Luminosity as a Reference SFR Indicator}\label{lirsfr}

\begin{figure*}
\begin{center}
\includegraphics[width=150mm]{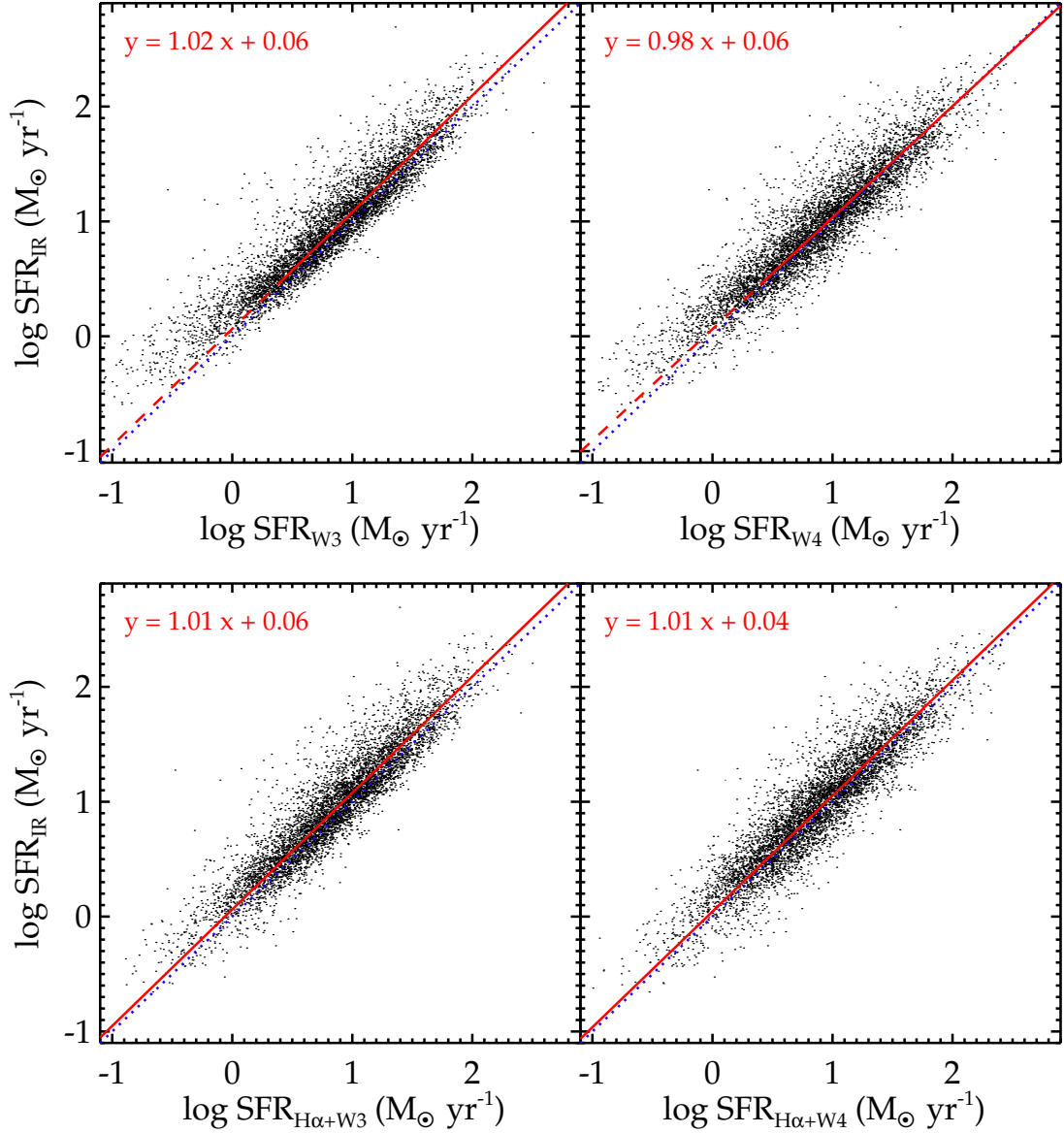} 
\end{center}
\caption{Comparison between SFRs based on total IR luminosity and on
  12 \micron{} luminosity (top left),
  22 \micron{} luminosity (top right),
  combination of H$\alpha$ and 12 \micron{} luminosities (bottom left), and
  combination of H$\alpha$ and 22 \micron{} luminosities (bottom right).
The red solid line is the best fit to the data, and 
  the blue dotted line is the one-to-one relation.
In the top panels,
  we use only galaxies with SFR $>$ 3 $M_{\sun}$ yr$^{-1}$ for the fit,
  and red dashed lines is an extension of the solid line.
}\label{fig-comp}
\end{figure*}

Figure \ref{fig-comp} shows the comparison of IR-based SFRs
  with other SFR estimates in this study (Equations 1--4).
The top panels
  show that SFR$_{\rm IR}$ and MIR-based SFRs correlate well
  for galaxies with SFR $\gtrsim$ 1 $M_{\odot}$ yr$^{-1}$.
We show the best fit to the data in each panel as a solid line. 
However, the slopes of the relations seem to change, 
  similar to the relations between SFR$_{\rm H\alpha}$ and MIR-based SFRs 
  (see top panels of Figure \ref{fig-mono}).
These results suggest that our calibration does not change much
  even if we use SFR$_{\rm IR}$ 
  as a reference indicator instead of SFR$_{\rm H\alpha}$.

The bottom panels show that SFR$_{\rm IR}$ and SFR$_{\rm H\alpha+MIR}$
  also agree well.
Moreover, the scatters in these 
  SFR$_{\rm H\alpha+MIR}$-SFR$_{\rm IR}$ relations ($\sim$0.25 dex) 
  are smaller than for the SFR$_{\rm H\alpha}$-SFR$_{\rm IR}$ relation 
  (0.31 dex; see the left panel of Figure \ref{fig-base}).
This suggests that the uncertainty in extinction correction 
  can be reduced if we use the combination of H$\alpha$ and MIR luminosities
  rather than the Balmer decrement.

\subsection{Limitations of Our Calibration}\label{limit}

In this study, we use SFRs
  converted from observed quantities (i.e., extinction-corrected 
  H$\alpha$ luminosities and total infrared luminosities)
  with the relations in \citet{ken98} as references for the calibration.
Therefore, our SFR recipes are 
  only valid under the assumptions for the SFR conversion
  relations (see \citealt{ken98} for details).
For example,
  the strong dependence of SFR$_{\rm W3}$/SFR$_{\rm H\alpha}$
  on metallicity in Figure \ref{fig-res1} could be
  affected by the assumption in the  SFR conversion relation.
When we convert H$\alpha$ luminosity into SFR$_{\rm H\alpha}$,
  we use a constant conversion factor of \citet{ken98}
  that is based on the assumption of solar metallicity.
However, the conversion factor could be smaller in metal-poor galaxies.
\citet{bri04} indeed showed that
  the Kennicutt conversion factor is a very good typical value,
  but can change by $\lesssim$0.4 dex depending on the metallicity.
However, 
  although we use SFR$_{\rm Opt}$ of Brinchmann et al. 
  that takes into account the variation of conversion factor,
  the metallicity dependence of SFR ratio still remains.
This trend is also confirmed by \citet{dom12}, who found that
  the SFR ratio between SFR$_{\rm IR}$
  and SFR$_{\rm H\alpha}$ still depends on metallicity
  even if they use the recipes of Brinchmann et al. to derive SFR$_{\rm H\alpha}$.
  
We also assume that all the MIR emission of galaxies
  is attributed to the current star formation.
Therefore, the SFRs based on our calibration 
  could overestimate the true SFRs of galaxies
  if the MIR emission is significantly contaminated by
  other components such as dust emission from AGN 
  (see next section for details).

The SFR indicators in this study are calibrated with 
  normal star-forming galaxies in the local universe.
Thus the SFR recipes may not be applicable to the galaxies
  not covered in this study.
For example, our sample does not contain the galaxies with very low SFRs
 (i.e., SFR $\lesssim$ 0.1 $M_{\odot}$ yr$^{-1}$)
 and with very high SFRs (i.e., SFR $\gtrsim$ 100 $M_{\odot}$ yr$^{-1}$).
It is also necessary to examine whether the SFR recipes determined
  with low-$z$ galaxies are still applicable to high-$z$ star-forming galaxies 
  (e.g., \citealt{mag10});
  high-$z$ galaxies may experience star formation 
  under the different physical conditions 
  from low-$z$ galaxies \citep[e.g.,][]{hwa10b,kaw11,elb11}. 
  
In heavily obscured galaxies,
  the H$\alpha$ luminosities could be underestimated
  if the Balmer decrement is used for the extinction correction.
This is because the correction is not meaningful at
  $V$-band optical depths $\gtrsim$ 5 \citep[e.g.,][]{vei99,mou06}.
This problem could be solved if we use H$\alpha$-based SFRs 
  with extinction corrections  based on emission lines at longer wavelengths 
  (e.g., Pa$\alpha$/H$\alpha$) or use other SFRs not severely affected by 
  dust emission (e.g., radio 20 cm continuum) \citep[see][]{ken09}.

The SFR recipes in this study are not applicable to individual \mbox{H\,{\sc ii}} regions 
  or star-forming complexes 
  because our calibration is based on the integrated properties of galaxies.
The comparison of spatially resolved H$\alpha$ and MIR images of star-forming galaxies
  suggests that there is a diffuse MIR emission
  other than MIR and H$\alpha$ emissions from point-like sources
  \citep[e.g.,][]{pre07,ken09,ken12}.
This diffuse component comes from the cool interstellar dust 
  (i.e., IR cirrus emission), and can contribute to the MIR emission of galaxies
  up to several tens of percent \citep[e.g.,][]{bel03,dal07}.
Therefore, the calibration of SFR indicators can be different 
  between galaxies and \mbox{H\,{\sc ii}} regions
  depending on the amount of diffuse MIR emission \citep[see][]{zhu08}.

\subsection{Contamination of AGN and Stellar Continuum to the MIR emission}\label{contam}

The MIR emission in star-forming galaxies
  is mainly dominated by dust continuum and PAH features, 
  associated with current star formation.
However, there could be other components contributing to the MIR emission: 
  dust emission from AGB stars and
  AGN, and remaining stellar continuum.
The AGB dust emission is already considered 
  in the bottom left panels of Figures \ref{fig-res1} and \ref{fig-res2};
  its contribution is insignificant
  in our sample of galaxies with SFR $\gtrsim$ 0.1 $M_{\odot}$ yr$^{-1}$.

The dust emission from AGN can be significant
  in IR luminous galaxies \citep[e.g.,][]{mul11,lee12}.
However, the AGN contribution in our sample is expected to be $\lesssim$ 10\%
  because we use only star-forming galaxies
  classified on the emission-line ratio diagram \citep[]{don12,lee12k}. 
Therefore, the effect of AGNs on our calibration of SFR indicators
  is very small.

The stellar continuum of galaxies peaks around the near-IR,
   but can remain even in the MIR.
If we assume that {\it WISE} 3.4 \micron{} flux density is dust-free,
  we can compute 
  the contribution of stellar continuum to the 12 \micron{} flux density
  by properly scaling 3.4 \micron{} flux density \citep[see][]{hel04,wu05}.
Using the {\it WISE} selected SDSS galaxies without optical emission lines 
  (i.e., no star formation and nuclear activity),
  we find that the scaling factor is 0.1
  from the ratio between {\it WISE} 12 and 3.4 \micron{} flux densities.
This scaling factor is comparable to the one in \citet[$\sim$0.15]{jar13}.
The corresponding contribution of stellar continuum 
  to the 12 $\micron$ flux density in our sample is then only a few percent.
We find that 
  the dependence of the ratio between SFR estimates on physical parameters
  and its scatter do not change even if we use stellar continuum subtracted 
  MIR luminosities for the calibration of SFR indicators.

\section{CONCLUSIONS}

We use {\it WISE} 22 \micron{} selected,
  star-forming galaxies at $0.01 < z < 0.3$ in the SDSS
  to calibrate the SFR indicators based on 
  12 and 22 \micron{} monochromatic luminosities
  and on the combination of MIR and H$\alpha$ luminosities.
We adopt extinction-corrected H$\alpha$ luminosities 
  and total IR luminosities as reference SFR indicators.
We then investigate how the calibration depends on physical parameters
  including metallicity, stellar mass, mean stellar age, 
  and dust extinction.
Our main results are:

\begin{enumerate}
\item
Both 12 and 22 \micron{} monochromatic luminosities 
  correlate well with reference SFR estimates (Equations 1 and 2).
However, these MIR luminosities, especially for 12 \micron, 
  tend to underestimate SFRs 
  of galaxies with SFR $\lesssim$ 1 $M_{\odot}$ yr$^{-1}$.
This discrepancy seems to mainly result from low metallicity effect.

\item
We confirm that the metallicity dependence of MIR-based SFRs
  can be reduced by using a linear combination observed H$\alpha$ and MIR luminosities
  ($L_{\rm H\alpha(obs)} + a~L_{\rm MIR}$).

\item
We find that the combination coefficient ($a$) increases with SFR.
To take into account this variation, we use a non-linear combination of 
  observed H$\alpha$ and MIR luminosities (Equations 3 and 4).
This method provides the best SFR estimates
  with small scatters and with little dependence on physical parameters.
\end{enumerate}

We confirm that {\it WISE} MIR
  monochromatic luminosities can be good SFR indicators of dusty galaxies,
  but suffer from metallicity bias.
To mitigate this metallicity dependence, 
  we applied the energy balance method 
  that combines (M)IR and UV/optical measurements
  to {\it WISE} data for the first time,
  providing robust SFR estimates
  with small scatters and with little dependence on physical parameters.
Our calibration is robust because it is 
  based on a large sample of galaxies with a wide SFR range 
  and on reliable reference SFR estimates;
  it is applicable to the galaxies 
  with 0.1--100 $M_{\odot}$ yr$^{-1}$.
The proposed SFR recipes will be useful for studying the star formation activity
  for a large sample of {\it WISE} selected galaxies.

\acknowledgments
We are grateful to an anonymous referee for his/her comments to improve the original manuscript.
We would like to thank Hyunjin Shim, Kwang-Il Seon and Margaret Geller for useful discussions.
J.C.L. and J.K. are members of the Dedicated Researchers for Extragalactic AstronoMy (DREAM)
in Korea Astronomy and Space Science Institute (KASI).
H.S.H. acknowledges the Smithsonian Institution for the support of his post-doctoral fellowship.
This publication makes use of data products from the {\it Wide-field Infrared Survey Explorer}, 
which is a joint project of the University of California, Los Angeles, 
and the Jet Propulsion Laboratory, California Institute of Technology, 
funded by the National Aeronautics and Space Administration.
Funding for the SDSS and SDSS-II has been provided by the Alfred P. Sloan Foundation, 
the Participating Institutions, the National Science Foundation, 
the U.S. Department of Energy, the National Aeronautics and Space Administration, 
the Japanese Monbukagakusho, the Max Planck Society, 
and the Higher Education Funding Council for England.
The SDSS Web Site is http://www.sdss.org/.
The SDSS is managed by the Astrophysical Research Consortium for the Participating Institutions. 
The Participating Institutions are the American Museum of Natural History, 
Astrophysical Institute Potsdam, University of Basel, University of Cambridge, 
Case Western Reserve University, University of Chicago, Drexel University, Fermilab, 
the Institute for Advanced Study, the Japan Participation Group, Johns Hopkins University, 
the Joint Institute for Nuclear Astrophysics, 
the Kavli Institute for Particle Astrophysics and Cosmology, 
the Korean Scientist Group, the Chinese Academy of Sciences (LAMOST), 
Los Alamos National Laboratory, the Max-Planck-Institute for Astronomy (MPIA), 
the Max-Planck-Institute for Astrophysics (MPA), New Mexico State University, 
Ohio State University, University of Pittsburgh, University of Portsmouth, 
Princeton University, the United States Naval Observatory, and the University of Washington.

\begin{deluxetable*}{cr@{~}lr@{~}lr@{~}lccr@{}l}
\tabletypesize{\footnotesize} 
\tablecaption{SFR calibrations based on monochromatic MIR luminosities\label{table1}}
\tablewidth{0pt}
\tablehead{
\colhead{Band} & \multicolumn{2}{c}{$m$} & \multicolumn{2}{c}{$~n$} & 
\multicolumn{2}{c}{$c$} & \colhead{$\sigma$} & \colhead{Ref.} & \multicolumn{2}{c}{SFR range}}
\startdata
{\it Spizter} ~8   & 1.09 & $\pm$ 0.06   &  $-$10.03 & $\pm$ 0.16   &  $-$9.20 & $\pm$ 0.19 &      & 1 &  0.2&--20  \\
{\it Spitzer} ~8   & 0.93 & $\pm$ 0.03   &   $-$8.59 & $\pm$ 0.08   &  $-$9.19 & $\pm$ 0.19 & 0.18 & 4 &  0.1&--30  \\
{\it AKARI} ~9     & 0.99 & $\pm$ 0.03   &   $-$8.84 & $\pm$ 0.32   &          &            & 0.18 & 6 &  0.2&--150 \\
{\it WISE} ~12     & 1.01 & $\pm$ 0.01   &   $-$8.85 & $\pm$ 0.01   &          &            &      & 7 &  0.1&--100 \\
{\it WISE} ~12     & 0.67 &              &   $-$6.45 &              &          &            &      & 8 & 0.01&--100 \\
{\it WISE} ~12     &      &              &           &              &  $-$9.13 & $\pm$ 0.03 & 0.03 & 9 & 0.03&--4   \\
{\it WISE} ~12     & 1.03 & $\pm$ 0.01   &   $-$9.02 & $\pm$ 0.02   &  $-$8.78 & $\pm$ 0.03 & 0.20 & 10 &   3&--100 \\
\cline{1-11}
{\it Spizter} ~24   & 0.89 & $\pm$ 0.06  &   $-$7.82 & $\pm$ 0.17   &  $-$8.81 & $\pm$ 0.19 &      & 1 &  0.2&--20  \\
{\it Spizter} ~24   & 0.87 &             &   $-$7.82 &              &          &            &      & 2 & 0.01&--50  \\
{\it Spizter} ~24   & 0.89 & $\pm$ 0.03  &   $-$7.97 & $\pm$ 0.97   &          &            & 0.30 & 3 & 0.01&--50  \\
{\it Spitzer} ~24   & 0.85 & $\pm$ 0.02  &   $-$7.47 & $\pm$ 0.06   &  $-$8.85 & $\pm$ 0.17 & 0.16 & 4 &  0.1&--30  \\
{\it Spitzer} ~24   &      &             &           &              &  $-$8.93 &            & 0.13 & 5 &    1&--10  \\
{\it AKARI} ~18     & 0.90 & $\pm$ 0.03  &   $-$7.85 & $\pm$ 0.30   &          &            & 0.20 & 6 &  0.2&--150 \\
{\it WISE} ~22      & 0.70 &             &   $-$6.75 &              &          &            &      & 8 & 0.01&--100 \\
{\it WISE} ~22      &      &             &           &              &  $-$8.94 & $\pm$ 0.01 & 0.01 & 9 & 0.03&--4   \\
{\it WISE} ~22      & 0.96 & $\pm$ 0.01  &   $-$8.37 & $\pm$ 0.02   &  $-$8.80 & $\pm$ 0.03 & 0.21 & 10 &   3&--100 \\
\enddata
\tablecomments{
$m$ and $n$ are the coefficients for the fit with log SFR$_{\rm MIR} = m$ log $L_{\rm MIR} + n$,
  and $c$ is for log SFR$_{\rm MIR} =$ log $L_{\rm MIR} + c$. 
$\sigma$ is the standard deviation of the fitting residuals.  
References are (1) \citet{wu05}, (2) \citet{alo06}, (3) \citet{cal07}, (4) \citet{zhu08}, (5) \citet{rie09}, 
  (6) \citet{yua11}, (7) \citet{don12}, (8) \citet{shi12}, (9) \citet{jar13}, and (10) This study.
Each calibration is derived in a given SFR range, and is matched to the Salpeter IMF.
Here the SFR and luminosity are expressed in units of $M_{\sun}$ yr$^{-1}$ and $L_{\sun}$, respectively.}
\end{deluxetable*}

\begin{deluxetable*}{cr@{~}lr@{~}lr@{~}lccr@{}l}
\tabletypesize{\footnotesize} 
\tablecaption{SFR calibrations based on combinations of H$\alpha$ and MIR luminosities\label{table2}}
\tablewidth{0pt}
\tablehead{
\colhead{Band} & \multicolumn{2}{c}{$a$} & \multicolumn{2}{c}{$i$} & 
\multicolumn{2}{c}{$j$} & \colhead{$\sigma$} & \colhead{Ref.} & \multicolumn{2}{c}{SFR range}} 
\startdata
{\it Spizter} ~8  & 0.010 &              &     1.12 & $\pm$ 0.01  &  $-$8.49 & $\pm$ 0.04  & 0.14 & 2 &   0.1&--30 \\  
{\it Spizter} ~8  & 0.011 & $\pm$ 0.003  &          &             &          &             & 0.11 & 3 & 0.004&--100\\  
{\it WISE} ~12    & 0.036 & $\pm$ 0.001  &     1.07 & $\pm$ 0.01  &  $-$8.12 & $\pm$ 0.02  & 0.17 & 4 &   0.1&--100\\  
\cline{1-11} 
{\it Spizter} ~24 & 0.031 & $\pm$ 0.006  &          &             &          &             & 0.30 & 1 &  0.01&--50 \\ 
{\it Spizter} ~24 & 0.022 &              &     1.04 & $\pm$ 0.01  &  $-$7.82 & $\pm$ 0.04  & 0.14 & 2 &   0.1&--30 \\ 
{\it Spizter} ~24 & 0.020 & $\pm$ 0.005  &          &             &          &             & 0.12 & 3 & 0.004&--100\\ 
{\it WISE} ~22    & 0.034 & $\pm$ 0.001  &     1.06 & $\pm$ 0.01  &  $-$8.04 & $\pm$ 0.02  & 0.18 & 4 &   0.1&--100\\ 
\enddata
\tablecomments{
$a$ is the coefficient for the fit with $L_{\rm H\alpha(corr)} = L_{\rm H\alpha(obs)} + a~L_{\rm MIR}$,
  and $i$ and $j$ are for log SFR$_{\rm H\alpha+MIR} = i$ log $(L_{\rm H\alpha(obs)} + a~L_{\rm MIR}) + j$.
$\sigma$ is the standard deviation of the fitting residuals.    
References are (1) \citet{cal07}, (2) \citet{zhu08}, (3) \citet{ken09}, and (4) This study.
Each calibration is derived in a given SFR range, and is matched to the Salpeter IMF.
Here the SFR and luminosity are expressed in units of $M_{\sun}$ yr$^{-1}$ and $L_{\sun}$, respectively.}
\end{deluxetable*}

\end{document}